\documentclass[]{JHEP3}
\usepackage{epsfig}
\title{Lattice QCD with mixed actions}

\author{UKQCD Collaboration, K.C. Bowler,
	B. Jo\'o,
	R.D. Kenway,
        C.M. Maynard
        and
        R.J. Tweedie\\
	School of Physics, The University of Edinburgh, 
        Edinburgh EH9 3JZ, UK}

\abstract{
We discuss some of the implications of simulating QCD when the action
used for the sea quarks is different from that used for the valence
quarks. We present exploratory results for the hadron mass spectrum
and pseudoscalar meson decay constants using improved staggered sea
quarks and HYP-smeared overlap valence quarks. We propose a method for
matching the valence quark mass to the sea quark mass and demonstrate
it on UKQCD clover data in the simpler case where the sea and valence
actions are the same.}

\preprint{Edinburgh 2004/27}

\begin{document}

\section{Introduction}

To predict phenomenological quantities from lattice QCD with high
precision requires the ability to simulate with light dynamical
quarks.  Ginsparg-Wilson fermions have the correct chiral and flavour
symmetries. However, they are computationally expensive compared to
improved staggered quarks. In the $N_f=2+1$ improved staggered
programme the square root of the fermion determinant is employed to
reduce the number of dynamical flavours from four to two for the up
and down quarks, and the fourth root is taken to reduce the number of
flavours from four to one for the strange
quark~\cite{Davies:2003ik}. Ensembles of gauge field configurations
are then generated with these fractional power determinants as weight
factors.  There is no known local action to which this model
corresponds. We define a mixed action as one where the action used to
generate the ensemble of gauge configurations, or sea quark action, is
different from the valence quark action used to determine hadronic
observables on those configurations. Current $N_f=2+1$ improved
staggered simulations have a mixed action because the four-flavour
staggered Dirac operator is used to generate the valence quark
propagators rather than a local operator equivalent to that used in
the ensemble weight. Unless a local operator can be found such that
\begin{equation}
 \det D_{\rm local} \equiv (\det\{ D_{\rm st}+m\})^{1/2}
\end{equation}
mixed actions are inevitable in the improved staggered programme. The
Chebyshev polynomial approximation to the square root of $(D_{\rm st}+m)$
is not the required operator as it has been shown to be
non-local~\cite{Bunk:2004br,Bunk:2004kf,Hart:2004sz}. That $(D_{\rm
st}+m)^{1/2}$ is non-local does not imply that $D_{\rm local}$ does not
exist, but serves as a warning, since the obvious candidate for such
an operator fails.

In the rest of this paper we assume that some $D_{\rm local}$ exists
so that the improved staggered ensembles are generated with an action
in the same universality class as QCD. We consider the case where the
valence quark action is manifestly different from that of the sea and
we choose the valence action that has the best chiral properties, that
is, overlap valence quarks on an improved staggered sea. In~\cite{Bar:2002nr} a
local Symanzik action and the corresponding low-energy chiral
effective Lagrangian are constructed for a general Ginsparg-Wilson
valence action with Wilson sea quarks. Some of their considerations
apply to more general mixed actions and, in particular, to overlap
valence quarks on a staggered quark sea~\cite{Bar:2004}.

Neuberger's overlap operator~\cite{Neuberger:1998my} is given by
\begin{equation}
D_{\rm ov}(\mu) = \frac{1}{2}\left[ 1 + \mu + \left(1 - \mu \right) 
 \gamma_5 \epsilon (H_{\rm W}(-\rho)) \right]
\end{equation}
where $H_{\rm W}$, is the Hermitian Wilson operator
\begin{equation}
H_{\rm W}(-\rho) = \gamma_5 D_{\rm W}(-\rho)
\end{equation}
with mass parameter $0 \le \rho \le 2$, and $\epsilon(H_{\rm W})$ is
the matrix sign function of $H_{\rm W}$.
The mass parameter $\mu$ is related to the bare quark mass $a m_q$ 
through
\begin{equation}
\label{eqn:mqO}
\mu = \frac{ a m_q }{2 \rho}\ 
\end{equation}
although we will ignore this below and write $D_{\rm ov}(m_{0})$.
The expectation value of some observable $\mathcal{O}$
in a model where the ensemble has been generated as $2+1$ flavours
of staggered quarks, with overlap valence quarks is
\begin{eqnarray}
  \langle {\mathcal O} \rangle &=& 
	\frac{1}{Z}\int {\mathcal D} U 
	\left(\det\left\{ D_{\rm st}[U] + m_{ud} \right\}\right)^{1/2}
	\left(\det\left\{ D_{\rm st}[U] + m_s \right\}\right)^{1/4}e^{-S_g[U]}
	\\\nonumber
  && \times{\mathcal O}\left[\frac{\delta}{\delta \bar{\eta}_i},
	\frac{\delta}{\delta {\eta}_i},U\right]
    \left.e^{-\bar{\eta}_i\left\{ D_{\rm ov}[U](m_i)\right\}^{-1}\eta_i}
	\right|_{\bar{\eta_i}=\eta_i=0},
\end{eqnarray}
where $U$ are the gauge fields, $Z$ is the partition function,
$\{\bar{\eta}_i,\eta_i\}$, $i=1,\cdots,N_f$, are the valence quark
sources and $S_g$ is the gauge action. The real parts the eigenvalues
of $D_{\rm ov}$ are positive and bounded from below by the valence
quark masses $m_i$, assuming $m_i>0$. The expectation values are equal
to those of a local field theory with action given by
\begin{eqnarray}
S &=& S_g[U] + 
   \sum_{l=ud} \bar{\chi}_l\left( D_{\rm local}[U] + m_{ud} \right) \chi_l
  + \bar{\chi}_s\left( D_{\rm local}[U] + m_{s} \right) \chi_s
	\\\nonumber
 &&  + \sum_i \left\{\bar{q}_i D_{\rm ov}[U](m_i) q_i + 
	\phi^+_i D_{\rm ov}[U](m_i) \phi_i \right\}
\end{eqnarray}
where the $\chi$ fields are the one-component staggered sea quark
fields, and the $q$ fields are the overlap valence quark fields. The
$\phi$ fields are pseudofermion sea fields introduced to cancel the
determinant of the overlap operator~\cite{Morel:1987xk}.  For practical purposes the model
can be regarded as having an exact $SU(N_{f}|N_{f})_{L}\otimes
SU(N_{f}|N_{f})_{R} \otimes U(1)_{V}$ symmetry when $m_i=0$ for $i=1,
\cdots ,N_f$~\cite{Sharpe:2001fh}. Restricting to transformations only in the valence
quark sector, the infinitesimal chiral rotation is given by
\begin{eqnarray}
  \delta q &=& i\epsilon\tau\gamma_5\left(1- \frac{1}{2}D_{\rm ov}\right)q
	\\\nonumber
  \delta \bar{q} 
	&=& i\epsilon\bar{q}\left(1- \frac{1}{2}D_{\rm ov}\right)\gamma_5\tau
\end{eqnarray}
and possesses the correct $U(1)_{A}$ anomaly and an index
theorem~\cite{Hasenfratz:1998ri,Luscher:1998pq} (for a review
see~\cite{Chandrasekharan:2004cn}).

For $N_{f} = 3$, this model is in the same universality class as QCD
when the sea and valence quark masses are matched. At non-zero lattice
spacing, the separate chiral symmetries for sea and valence quarks
ensure that the lightest pseudoscalar meson mass vanishes at
$m_{\rm val}=m_{\rm sea}=0$. This implies that the bare
quark masses are related by
\begin{equation}
\label{eqn:mqMatch}
  m_{\rm val} = \zeta(a) m_{\rm sea}
\end{equation}
where $\zeta\rightarrow 1$ as $a\rightarrow 0$. To date $N_f=2+1$
simulations with staggered valence and fractional determinants of the
staggered sea~\cite{Davies:2003ik} have set $\zeta(a)=1$. However, it
is not obvious that this is the appropriate matching condition for
overlap valence quarks on a staggered sea (or, for that matter, for
staggered valence quarks).

\subsection{Matching the quark masses}
To match the sea and valence quark masses to their experimental values
one would have to find an experimentally known hadronic state whose
mass depends strongly on the sea quark mass. In principle, the
$\eta^\prime$ is one such hadron. The sea quark mass could be tuned
until the $\eta^\prime$ has the correct experimental mass, whilst
tuning the valence quark mass of the flavour non-singlet pseudoscalar
meson to the pion. In practice, this is rather difficult, as the
$\eta^\prime$ requires very high statistics calculations. An
alternative would be to relate the bare sea and valence quark masses
to each other via equation (\ref{eqn:mqMatch}), and then tune the
flavour non-singlet mesons to their experimental values in the usual
way.

When the sea quark mass is infinite, {\em i.e.} quenched, then
Bardeen~{\em et al.}~\cite{Bardeen:2001jm,Bardeen:2003qz} have demonstrated
numerically that the model violates unitarity. This has also been observed numerically
in partially quenched two flavour QCD and demonstrated in partially quenched chiral
perturbation theory by~\cite{Prelovsek:2004jp}.
We perform a similar analysis and show the same unitarity 
violation occurs when $m_{\rm val} < m_{\rm sea}$. Our results suggest
a criterion for matching the sea and valence quark masses.  The quark
masses can be tuned by varying the valence quark mass to see when
these partially quenched pathologies appear for a given sea quark
mass. This determines when the valence quark is lighter than the sea
quark.

Bardeen~{\em et al.}~\cite{Bardeen:2001jm,Bardeen:2003qz} show that the scalar correlator,
\begin{equation}
  C_{SS}(t) = \sum_{\vec{x}} e^{-\vec{p}\cdot\vec{x}} 
	\langle \bar{\psi}(x)\psi(x) \bar{\psi}(0)\psi(0)\rangle
\end{equation}
is sensitive to this quenched pathology, because it couples to an
$\eta^\prime - \pi$ intermediate state. Shown in figure
\ref{fig:hairPin} are two of the diagrams which contribute to the
$\eta^{\prime}$ propagator.  Diagram a), the ``hairpin'', has a
negative coefficient. In full QCD, diagram b), with a series of vacuum
bubbles, cancels the effect of the hairpin diagram, so there is no
negative contribution. In quenched QCD, only the hairpin diagram
contributes, so the intermediate $\eta^{\prime}-\pi$ state couples
with a negative spectral weight. This gives the scalar correlator a
negative value.

\FIGURE{
  \epsfig{file=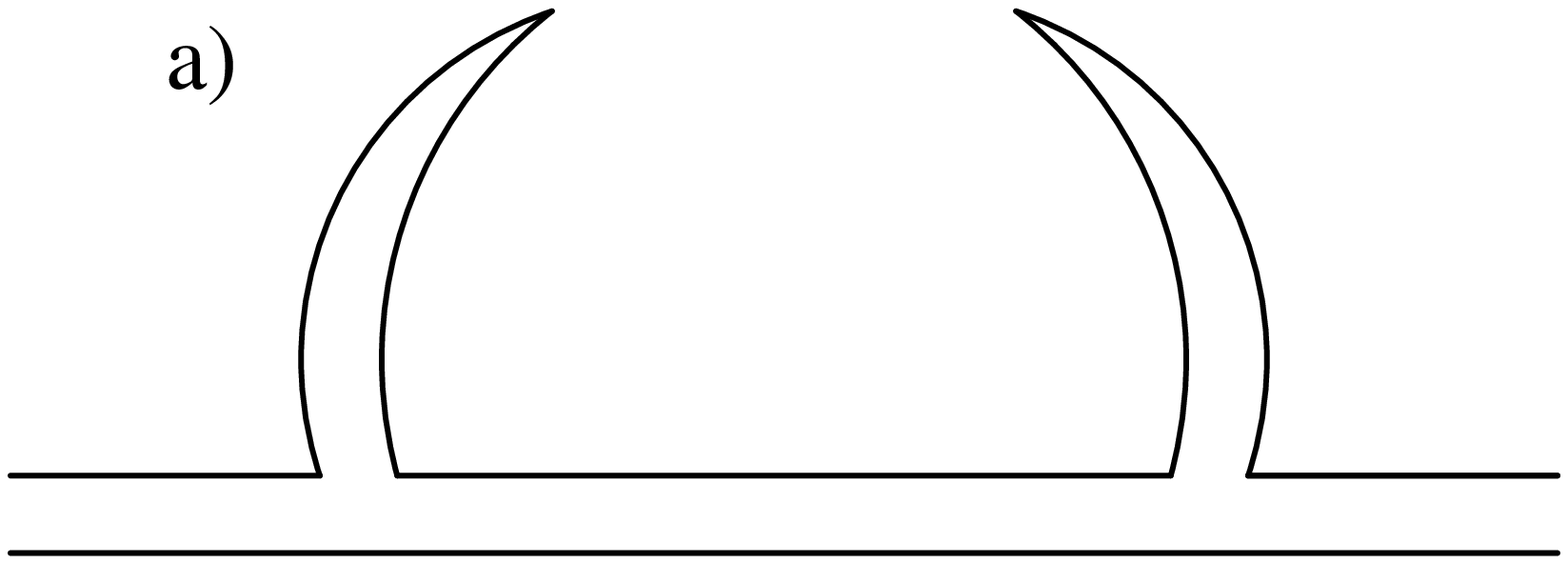,width=5cm}
  \hspace{1cm}
  \epsfig{file=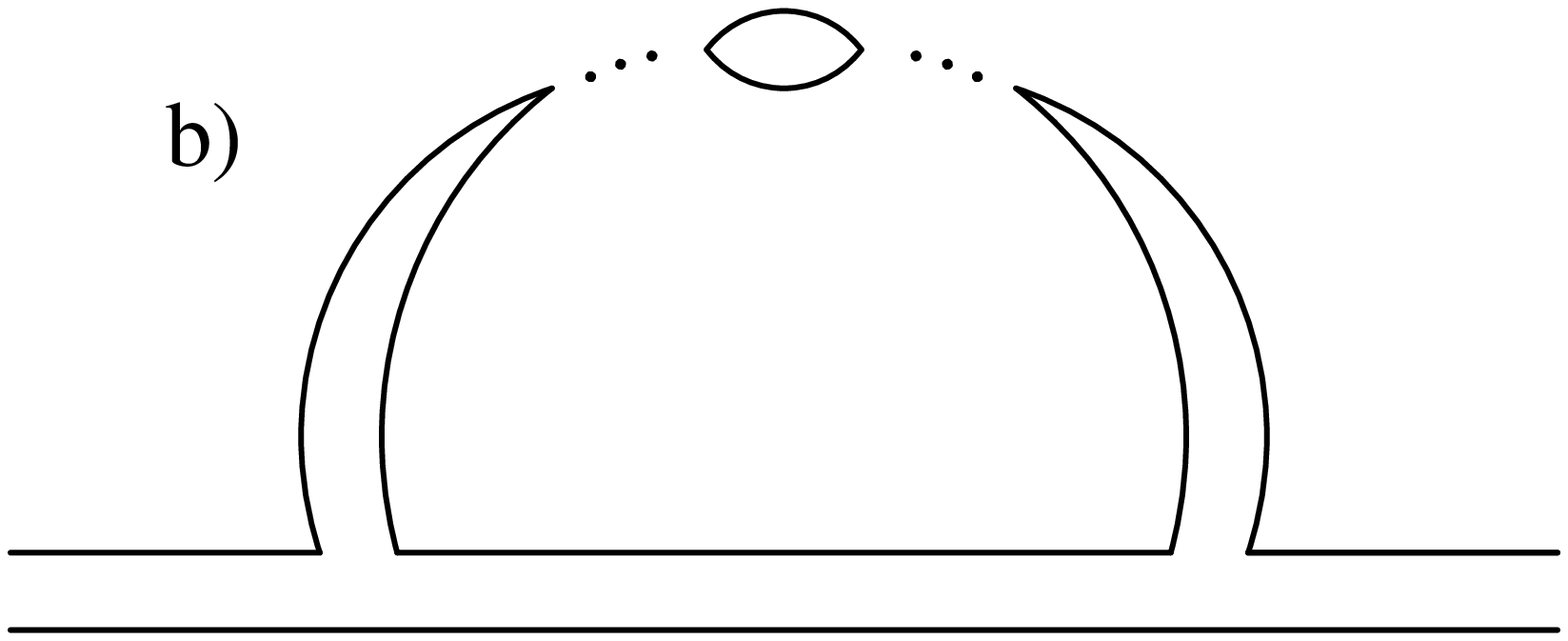,width=5cm}
  \caption{Quark-flow diagrams contributing to the $\eta^{\prime}$ propagator.}
  \label{fig:hairPin}
}

In partially quenched QCD the situation is more complicated. The
bubble in diagrams b) depends only on the sea quark mass, whereas the
connected quark-flow lines depend only on the valence quark
mass. Heuristically at least, the size of the contribution from
diagrams b) can been thought of in the following way.  When the sea
quark mass is smaller than the valence quark mass, diagrams b) have a
larger positive contribution than the negative contribution from
diagram a). When the sea quark mass is heavier than the valence quark
mass, diagram b) has a smaller contribution than a), which means the
$\eta^{\prime}-\pi$ intermediate state couples to the scalar
correlator with a negative weight. By monitoring the sign of the
scalar correlator as the valence quark mass is varied it should be
possible to match the sea and valence quark masses.

To demonstrate this method, we examined the scalar correlator on the
UKQCD $N_f=2$ clover data sets~\cite{Allton:2001sk,Allton:2004qq},
where the sea and valence quarks have the same action. So whether the
valence quark mass is heavier or lighter than the sea is known. This
data was generated with the Wilson plaquette gauge action and the
clover quark action, where the coefficient of the
Sheikholeslami-Wohlert term~\cite{Sheikholeslami:1985ij} was
determined non-perturbatively~\cite{Jansen:1995ck}.  For all data sets
$\beta=5.2$ and the volume is $L^3\times T=16^3\times 32$.  The values
of the hopping parameter for the sea and valence quark masses, and the
number of configurations are shown in Table~\ref{tab:cloverData}.

The relatively poor signal-to-noise ratio for the scalar correlator 
implies the need for a large number of configurations. To improve
the statistical resolution, we used a ratio of correlation functions, 
as the statistical fluctuations are correlated. In particular,
we considered the ratio
\begin{equation}
  R(t) = \frac{C_{PP}(t)-C_{SS}(t)}{C_{PP}(t)}
\end{equation}
where $PP$ denotes the pseudoscalar correlator. At sufficiently large times the ground states
will dominate and $R$ is then
\begin{equation}
\label{eqn:R}
 R(t)=
	1 - \left(\frac{A_{SS}}{A_{PP}}\right)
	\left(\frac{e^{-m_St} + e^{-m_S(T-t)}}
	{e^{-m_Pt} + e^{-m_P(T-t)}}\right).
\end{equation}
At the mid-point of the lattice 
\begin{equation}
R(T/2) = 1 - \frac{A_{SS}}{A_{PP}}e^{-\Delta m T/2}
\end{equation}
where $\Delta m=m_S-m_P$ is the mass splitting between the scalar and
pseudoscalar states.  For a large enough lattice time extent, $T$,
this ratio tends to unity at the mid-point.

However, when the valence quark mass is lighter than the sea quark
mass, the $\eta^\prime-\pi$ state couples to the scalar correlator
with a negative weight. Thus, a signal for the valence quark mass
being lighter than the sea is $R>1$.  Figure~\ref{fig:ratio1} shows
the ratio for different sea and valence quark masses. The open circle
and filled square both have the sea and valence quark masses equal,
and $R$ tends to unity at the mid point of the lattice. For the filled circles, $R>1$
as $t\approx T/2$ at the $2\sigma$ level, a signal for partial quenching,
and indeed this data set has $m_{\rm val}<m_{\rm sea}$. This effect is
clearly dependent on the sea quark mass, as the open and filled
circles both have the same valence quark mass.

\FIGURE{
\epsfig{file=RB52K3500.eps,width=10cm}
\caption{$R(t)$ in equation (\ref{eqn:R}) versus Euclidean time, $t$.}
\label{fig:ratio1}
}

\TABLE{
\caption{UKQCD dynamical clover ($N_f=2$) data sets for $\beta=5.2$. }
\label{tab:cloverData}
\begin{tabular}{ccccr@{.}l}
\hline
$\kappa_{\rm sea}$ & $\kappa_{\rm val}$ &Nconfig& 
	$m_{\rm val}:m_{\rm sea}$ &
	\multicolumn{2}{c}{$A_{SS}/A_{PP}$}\\\hline
		  & 0.13400             & 202 &$>$ &$0$&$6(1)$\\		
                  & 0.13450	       & 202 &$>$ &
	\multicolumn{2}{l}{$190(50)$}\\
\raisebox{1.5ex}[0pt]{0.13500} & 0.13500 & 202 &$=$ &$0$&$0(2)$\\
                  & 0.13550	       & 202 &$<$ &$-0$&$00015(5)$\\ \hline
                  & 0.13500	       & 208 &$>$ &$1$&$2(2)$\\
\raisebox{1.5ex}[0pt]{0.13550} & 0.13550 & 208 &$=$ &$0$&$0(1)$\\ \hline
                  & 0.13550	       & 141 &$>$ &$1$&$5(5)$\\
0.13565            & 0.13565	       & 141 &$=$ &$5$&$0(10)$\\
                  & 0.13580	       & 141 &$<$ &$-0$&$00014(14)$\\\hline
                  & 0.13565	       & 137 &$>$ &$0$&$06(4)$\\ 
0.13580            & 0.13580	       & 137 &$=$ &$0$&$0(1)$\\ 
                  & 0.13595	       & 137 &$<$ &$-0$&$003(2)$\\ \hline 
\end{tabular}
}

Also shown in Table~\ref{tab:cloverData} is the result of fitting
equation (\ref{eqn:R}) to the data. Clearly the ratio $A_{SS}/A_{PP}$
is not a very well-determined quantity. However, it seems clear that
this ratio being negative is a signal at the $1-2\sigma$ level that
the data is partially quenched. Figure~\ref{fig:ratio2} shows both the
scalar correlator and the ratio (\ref{eqn:R}). At lighter quark mass
and with fewer configurations, the fit results become rather dependent
on the fit range chosen, but combining the fit information and
examining these plots, it is clear that for $\kappa_{\rm sea}=0.13565$,
$\kappa_{\rm val}=0.13580$ there is a signal for the negative weight
state, and for $\kappa_{\rm val}=0.13565$ this signal is absent.

\FIGURE{
\epsfig{file=RatiosKS0.13565.eps,width=10cm}
\caption{$R(t)$ and $C_{SS}(t)$ versus $t$ for $\kappa_{\rm sea}=0.13565$.}
\label{fig:ratio2}
}

A precise matching of the sea and valence quark masses will be
difficult to achieve, because the signal for the scalar ground state
at large times for light quarks seems to disappear into the
noise. When the valence quark mass is lighter than the sea, the signal
for the negative weight $\eta^\prime - \pi$ state is fairly
strong. However, our results suggest that it is possible, in principle,
(equivalently with very high statistics) to match the valence and sea
quark masses. This is necessary to make sense of simulations with
mixed actions when at least one of the sea or valence quark masses is
out-with the chiral regime and matching to chiral perturbation theory
is problematic.

\section{Overlap valence quarks on a staggered sea}

We have performed an exploratory study of overlap valence quarks on
the MILC $N_f=2+1$ improved staggered
configurations~\cite{Bernard:2001av}. We measure the simplest states
of the light hadron spectrum, mesons and baryons, and the pseudoscalar
decay matrix element for both light and heavy-light states. Due to a
lack of computational resources, the number of configurations analysed
was small. This prevented any realistic attempt at matching the sea
and valence quark masses as described in the previous section. Whilst
the results presented below, in figures 5 to 10, appear encouraging,
the low statistics means they must be regarded as purely illustrative of
the effectiveness of a mixed action approach.

\subsection{Smearing}
The overlap operator is only local for gauge configurations which are
``smooth enough''~\cite{Hernandez:1998et}. The MILC configurations we
used have a lattice spacing of $a\sim0.125$ fm and so are relatively
coarse. Smoothing the gauge configurations should improve the
localisation of the overlap operator. Moreover, smoothing the gauge
fields by ``HYP-smearing''~\cite{Hasenfratz:2001hp} can improve the
spectral properties of the Wilson-Dirac
operator~\cite{DeGrand:2002vu}, which reduces the amount of
computation required in the solver used to apply $\epsilon (H_{\rm
W})$. Indeed, HYP-smearing the gauge configuration does speed up the
inversions. Furthermore, the low-lying eigenvalues of the staggered
operator ``mimic'' the eigenvalue spectrum of the overlap operator
when the configurations are smoothed in this
way~\cite{Durr:2003xs,Durr:2004as,Durr:2004rz} suggesting that a
smoothly behaved matching condition may exist for light quark masses.

To examine the effect of multiple iterations of HYP-smearing, we studied
the quark-antiquark potential on 624 quenched UKQCD configurations at
$\beta=5.93$ with a volume of $16^{3}\times32$. The smearing parameters
used were $\alpha_1=0.75$, $\alpha_2=0.60$, and $\alpha_3=0.35$~\cite{Hasenfratz:2001hp}.
Planar Wilson loops were used to extract the
quark-antiquark potential, which was fitted to
\begin{equation}
  V(r)= V_0 + \sigma r - \frac{\kappa}{r}
\end{equation}
Figure~\ref{fig:sigma} shows the effect of multiple iterations of
HYP-smearing on the string tension, $\sigma$.
\FIGURE{
\epsfig{file=sigma.eps,width=10cm}
\caption{The effect of HYP-smearing on the long-range potential as 
	measured by $\sigma$.}
\label{fig:sigma}
} 
Repeated HYP-smearing quickly altered the short-distance behaviour, while the
medium-to-long distance behaviour remained relatively unchanged for a
small number ($\lesssim 3$) of iterations. The effect on the potential
of smoothing configurations has been studied many times before,
following the work of Teper~\cite{Teper:1991un}, and recently an
extensive study for different actions and different smearings has been carried
out~\cite{Durr:2004xu}. Our limited study agrees with these previous
results. We conclude for $N\le 3$ the effect of smearing does not significantly  alter the
long range potential and thus the spectrum.

\subsection{The light hadron spectrum}

The overlap propagator calculations were performed on ten
configurations from each of two ensembles produced by the MILC
collaboration~\cite{Bernard:2001av}. One ensemble has $am_s=0.05$,
$am_l=0.03$ and the other has $am_s=0.05$, $am_l=0.02$.  Both have a
lattice spacing $a \simeq 0.125$ fm and linear size $L \simeq 2.5$
fm. Three iterations of HYP-smearing were applied to each
configuration. The overlap operator from the SZIN code \cite{SZIN} was
then used to calculate propagators. These were created with seven
different valence quark masses using the overlap multi-mass solver:
four light and three heavy \cite{Dong:2001fm}.  Some of these results
have been previously reported in~\cite{Bowler:2004kv}.

We performed simultaneous fits to three different correlators in order
to extract the pseudoscalar meson mass (see figure
\ref{effective_mass}).
\FIGURE{
\epsfig{file=LLsimEffMass.eps,width=10cm}
\caption{Pseudoscalar meson effective mass and simultaneous uncorrelated 
fit to three correlators ($P=\bar{q}\gamma_5 q,\
A_4=\bar{q}\gamma_4\gamma_5 q$). The squares and diamonds are slightly
offset horizontally for clarity.}
\label{effective_mass}
} The fluctuations in the effective mass are larger than the apparent
statistical errors, but this is probably due to underestimation of the
variance on ten configurations.

A partially quenched analysis was carried out, that is the sea quark
mass was held fixed whilst varying the valence quark mass. Since we
had multiple input valence masses, non-degenerate meson correlators
could be constructed. Shown in figure \ref{fig:mpiVsmq} is the
two-dimensional fit performed to $(aM_{PS})^{2}$ versus valence
masses $m_{q_{1}}$ and $m_{q_{2}}$, which allowed evaluation of the
average $u$ and $d$ quark mass, $\hat{m}$, from
\begin{equation}
M_{\pi}^{2} = B\left(m_{q_{1}}+m_{q_{2}}\right) + A = 2B\hat{m} + A 
\end{equation}
where $M_{\pi}$ is the physical pion mass. The chiral symmetry of the
operator should make $(aM_{PS})^2$ vanish at zero quark mass. We do not
constrain the fit to satisfy this condition, but within the limited statistics
the parameter $A$ is consistent with zero.
This in turn allowed us to evaluate the strange quark mass, $m_{s}$, from
\begin{equation}
M_{K}^{2} = B\left(m_{s}+\hat{m}\right) + A 
\end{equation}
where $M_{K}$ is the physical kaon mass.

\FIGURE{
\epsfig{file=mpisqr_vs_mqave_m020_m030.eps,width=10cm}
\caption{The square of the pseudoscalar meson mass vs bare overlap quark mass.}
\label{fig:mpiVsmq} 
}

We also determined the masses of the nucleon and delta baryon. The signal
for the nucleon mass is very clean. Figure~\ref{fig:effMN}
shows the effective mass of the nucleon for the two operators
\begin{eqnarray}
  N_1(x)&=& \varepsilon_{ijk}(\psi^T_i C \gamma_5 \psi_j) \psi_k\\\nonumber
  N_2(x)&=& \varepsilon_{ijk}(\psi^T_i C \gamma_4\gamma_5 \psi_j) \psi_k.
\end{eqnarray}
It is remarkable that we can see a signal for the negative parity partner of
the nucleon on as few as ten configurations. This suggests that, despite their relative
cost per propagator compared with staggered quarks, overlap valence quarks maybe the
most cost effective way to extract precision light baryon physics from improved staggered
configurations.

\FIGURE{
\epsfig{file=nucleon_GW0.2_Sea0.3.eps,width=10cm}
\caption{Nucleon effective mass for the heavier sea quarks, with 
	 $am_{\rm q}=0.056$ (equation \ref{eqn:mqO}). The square
	 symbols show the negative parity excitation.}
\label{fig:effMN}
}

Figure \ref{fig:NandD} shows the nucleon (upper plot) and decuplet
(lower plot) masses versus the pseudoscalar meson mass squared. The
lines are uncorrelated linear fits to the data. The values
calculated by the MILC collaboration
\cite{Bernard:2001av} on their corresponding full ensembles are shown
by open symbols. Both the nucleon and decuplet baryon masses from the
overlap operator are significantly lower although, {\em a priori}, we
don't know how to match the horizontal scales.  The cut-off effects
for the different formalisms will be different and, unless the
matching function in equation (\ref{eqn:mqMatch}) is very different
from one, this suggests that the cut-off effects for the overlap
baryons are be smaller.  The nucleon mass shows some sea quark mass
dependence, but the decuplet mass shows no variation. With ten
configurations and relatively heavy sea quark masses, any trend is
hard to spot.  In the lower plot, the vertical dashed-dotted line
shows the estimate of the $\eta_{s\bar{s}}$ mass squared, as measured
by the overlap operator on these configurations. The horizontal dotted
line is the physical $\Omega^-$ mass in lattice units. Within large
statistical uncertainties, this determination of the $\Omega^-$ mass
at fixed lattice spacing agrees with the experimental value. Again,
this may suggest that cut-off effects with overlap fermions are
smaller than with staggered fermions, but, with data at only one
lattice spacing, this remains speculation.

\FIGURE{
\epsfig{file=NandDvsMPSsq.eps,height=10cm}
\caption{The nucleon and decuplet baryon mass versus $(aM_{PS})^{2}$ for the two ensembles. 
         The open symbols show the baryon masses measured on the
         full ensemble with staggered valence quarks~\cite{Bernard:2001av}.
	 The lower plot shows the physical $\Omega^-$ mass in lattice units, where
         the lattice spacing is set by $r_0$ taken from the reference above.}
\label{fig:NandD}
}

The pseudoscalar decay constant, $f_{PS}$, is defined as
\begin{equation}\label{fPS}
f_{PS} = \frac{Z_{A}\langle0|A_{4}|PS\rangle}{M_{PS}} .
\end{equation}
We obtain $Z_{A}$ from the axial Ward identity
\begin{equation}
Z_{A}\langle\partial_{\mu}A_{\mu} \mathcal{O} \rangle = 2m_{q}\langle
P \mathcal{O} \rangle
\end{equation}
which we can express in terms of the pseudoscalar correlator,
$C_{PP}$, and the pseudoscalar axial correlator,
$C_{PA_{4}}$. $\langle0|A|PS\rangle$ cancels in equation (\ref{fPS}) and
hence we require only $C_{PP}$ to compute a renormalised
$f_{PS}$. Once again, we performed a 2-d linear fit to the light
non-degenerate pseudoscalars to calculate $f_{PS}$ (see figure
\ref{Pseudoscalardecay}) and extracted the ratio of
$f_{K}/f_{\pi}$ (see table \ref{table1}). The value increases
slightly with decreasing light sea quark mass in the right direction
to agree with experiment. This is also evident from the slight change
of the gradient with sea quark mass in figure \ref{Pseudoscalardecay}.
\FIGURE{
\epsfig{file=Pseudoscalardecay.eps,height=10cm}
\caption{$f_{PS}$ versus $M_{PS}^{2}$ for the two ensembles.}
\label{Pseudoscalardecay}
}

\TABLE{
\caption{Pseudoscalar meson decay constants.}
\label{table1}
\newcommand{\m}{\hphantom{$-$}}
\newcommand{\cc}[1]{\multicolumn{1}{c}{#1}}
\renewcommand{\tabcolsep}{1pc} 
\renewcommand{\arraystretch}{1.2} 
\begin{tabular}{@{}lll}
\hline
Sea Quarks            & $f_{K}/f_{\pi}$ & $f_{Ds}$   (MeV)  \\
\hline
$am_{\rm{sea}}$ = 0.03/0.05     	  & \m1.03(3) & \m226(14)\\
$am_{\rm{sea}}$ = 0.02/0.05                & \m1.08(4) & \m232(11)\\
\hline
Experiment \cite{Eidelman:2004wy}	    & \m1.22(1) & \m 266(32)\\
\hline
\end{tabular}\\[2pt]
}

\subsection{Charm Physics}

Heavy quark propagators essentially come for free in the overlap
propagator calculation through the use of a multi-mass
solver. However, lattice artefacts are ${\mathcal{O}}(am_{q})^{2}$ and the
heaviest input valence quark mass used is $am_{q} = 0.84$, so
$(am_{q})^{2} \sim 0.7$. With the lattice spacing of $a^{-1} \sim
1.5$GeV, the calculation is at best on the limit of simulating
charm. Due to the rapid decay in Euclidean time, we require double
precision. However, this does not slow the solver down appreciably, as
we need substantially fewer re-orthogonalisations against the projected
eigenvectors of $H_{W}$ in the linear solver than in single precision.

These heavy quark propagators were used to calculate $f_{Ds}$ (see
table \ref{table1}). The value of $f_{Ds}$ increases with decreasing
light sea quark mass, in the direction of the experimental value, as
can be seen from the change of gradients in figure \ref{fDs}.
\FIGURE{
\epsfig{file=fDs.eps,width=10cm}
\caption{$f_{Hs}$ vs inverse heavy-strange pseudoscalar meson mass.}
\label{fDs}
}

The short distance behaviour of the potential has been altered by
repeated smearing. As heavy quarks in quarkonium feel the short distance potential,
this repeated smearing may be a source of worry. Indeed, examining
the heavy-heavy correlator for the heaviest quark mass we do
not see the effective mass reaching a plateau. It might be expected that a
heavy-light state feels the effect of the short distance potential
less. Indeed the effective mass for the heavy-light
correlator reaches a plateau. This suggests that the heavy-light
states are not suffering so much from the modified short-distance behaviour.
In future work we anticipate using fewer iterations of smearing.

\section{Conclusions}
While staggered quarks offer the most cost effective way of simulating
light dynamical quarks today, they require us to use a mixed action
formulation of QCD. Outside the chiral regime of both valence and sea
quarks, it is necessary to implement a matching procedure for the
quark masses for the model to be in the same universality class as
QCD. (Within the chiral regime, the partially quenched results may be
matched to chiral perturbation theory and thence to QCD low energy
constants.) Indeed, we show numerically that the partially quenched
theory with $m_{\rm val} < m_{\rm sea}$ has similar negative metric
pathologies to those observed by Bardeen {\em et al.} in quenched
QCD. In principle, this observation provides a matching condition, but,
just like the alternative approach of determining $m_{\rm sea}$ by
matching a flavour singlet quantity to experiment, suffers from poor
signal-to-noise in practice. Despite these practical problems with matching,
we obtain encouragingly good signals for flavour non-singlet hadron masses and
decay constants using overlap valence quarks on a staggered sea quark 
ensemble. The potential gain from the simplicity of valence quarks with
the correct flavour and chiral symmetries, together with the clean statistical
signals, particularly  for baryons, is good motivation for trying to
improve on our exploratory attempts to match valence and sea quark masses.

\acknowledgments
We thank S.~Sharpe for useful discussions.

\providecommand{\href}[2]{#2}\begingroup\raggedright\endgroup

\end{document}